\documentclass[12pt]{article}
\usepackage{amsthm}
\usepackage{graphicx}
\usepackage{enumerate}
\usepackage{natbib, threeparttable}
\usepackage{url} % not crucial - just used below for the URL 
\usepackage{setspace, bm} 

\usepackage{amsmath}
\usepackage{enumerate}

\usepackage{multirow}
\usepackage{verbatim}
\usepackage[font=footnotesize]{subcaption}
\usepackage{amssymb}
\usepackage{tikz}
\usepackage{booktabs}
\usepackage[normalem]{ulem} 

\usepackage{changepage}

\usepackage{enumerate}
\usepackage{url} % not crucial - just used below for the URL 
\usepackage{setspace, bm} 

\usepackage{amsmath}

\usepackage{multirow}
\usepackage{verbatim}
\usepackage[font=footnotesize]{subcaption}
\usepackage{amssymb}
\usepackage{tikz}
\usepackage{booktabs}
\usepackage[normalem]{ulem} 

\usepackage{changepage}

\usepackage{setspace}

\usepackage{upgreek}

\usepackage{enumitem, xcolor}

\newcommand{\blind}{1}

\theoremstyle{plain}

\newtheorem{theorem}{Theorem}[section]

\newtheorem{corollary}{Corollary}[theorem]
\theoremstyle{remark}

\newcommand{\sg}[1]{\textcolor{black}{#1}}

\begin{document}

\def\spacingset#1{\renewcommand{\baselinestretch}%
{#1}\small\normalsize} \spacingset{1}

%%%%%%%%%%%%%%%%%%%%%%%%%%%%%%%%%%%%%%%%%%%%%%%%%%%%%%%%%%%%%%%%%%%%%%%%%%%%%%

\if1\blind
{
  \title{\bf Causal Meta-Analysis by Integrating Multiple Observational Studies with Multivariate Outcomes}
   \author{Subharup Guha\thanks{ This work was supported by the National Science Foundation and National Institutes of Health under award
DMS-1854003 to SG, award CA249096  to YL,  and award CA269398 and CA209414 to SG and YL. 
}
\hspace{.2cm}\\
    Department of Biostatistics,  University of Florida\\
    and \\
    Yi Li\\
    Department of Biostatistics, University of Michigan}
  \maketitle
} \fi

\if0\blind
{
  \bigskip
  \bigskip
  \bigskip
  \begin{center}
    {\LARGE\bf   \title{\bf Causal Meta-Analysis by Integrating Multiple Observational Studies with Multivariate Outcomes}}
\end{center}
  \medskip
} \fi

\bigskip

\begin{abstract}
Integrating multiple observational studies to make unconfounded causal or descriptive comparisons of group potential outcomes in a large natural population is challenging. Moreover, retrospective cohorts, being convenience samples, are usually unrepresentative of the natural population of interest and have groups with unbalanced covariates.  We propose a general covariate-balancing framework based on pseudo-populations that  extends established weighting methods to the meta-analysis of multiple retrospective cohorts with multiple groups. Additionally, by maximizing  the effective sample sizes of the cohorts, we propose a \underline{FLEX}ible, \underline{O}ptimized, and \underline{R}ealistic (FLEXOR)  weighting method appropriate for integrative analyses. We develop new weighted estimators for unconfounded inferences on wide-ranging population-level  features and estimands  relevant to  group comparisons of quantitative, categorical, or multivariate outcomes. Asymptotic properties of these estimators are examined. Through simulation studies and  meta-analyses of TCGA datasets, we demonstrate the versatility and reliability of the proposed weighting strategy, especially for the FLEXOR pseudo-population.
\end{abstract}

%  Please place your key words in alphabetical order, separated
%  by semicolons, with the first letter of the first word capitalized,
%  and a period at the end of the list.
%

\textbf{KEYWORDS}:
FLEXOR; 
 Pseudo-population; Retrospective cohort; Unconfounded comparison; Weighting.

%  As usual, the \maketitle command creates the title and author/affiliations
%  display 

\maketitle

%  If you are using the referee option, a new page, numbered page 1, will
%  start after the summary and keywords.  The page numbers thus count the
%  number of pages of your manuscript in the preferred submission style.
%  Remember, ``Normally, regular papers exceeding 25 pages and Reader Reaction 
%  papers exceeding 12 pages in (the preferred style) will be returned to 
%  the authors without review. The page limit includes acknowledgements, 
%  references, and appendices, but not tables and figures. The page count does 
%  not include the title page and abstract. A maximum of six (6) tables or 
%  figures combined is often required.''

%  You may now place the substance of your manuscript here.  Please use
%  the \section, \subsection, etc commands as described in the user guide.
%  Please use \label and \ref commands to cross-reference sections, equations,
%  tables, figures, etc.
%
%  Please DO NOT attempt to reformat the style of equation numbering!
%  For that matter, please do not attempt to redefine anything!

\section{Introduction} \label{S:intro}

The study of differential patterns of oncogene expression levels across cancer subtypes has aroused great interest because it unveils new tumorigenesis mechanisms    and can improve  cancer screening  and treatment \citep{kumar2020oncogenic}. In a multi-site breast cancer study conducted  at seven medical centers, including, for example, 
Memorial Sloan Kettering, Mayo Clinic, and University of Pittsburgh, 
the goal was to compare  the mRNA expression levels of  eight targeted breast cancer genes,  namely, COL9A3, CXCL12, IGF1, ITGA11, IVL, LEF1,  PRB2,  and SMR3B  \citep[e.g.,][]{christopoulos2015role} in  the disease subtypes infiltrating ductal carcinoma (IDC) and infiltrating lobular carcinoma (ILC),  which account  for  nearly 80\% and 10\%  of breast cancer cases in the United States \citep{IDC,ILC}.
  The  data 
    reposited at The Cancer Genome Atlas (TCGA) portal \citep{GDC} include      demographic, clinicopathological, and biomarker measurements;  some study-specific  attributes are summarized in  Supplementary Materials. Each breast cancer patient's outcome is  a  vector   of   mRNA expression measurements for these eight targeted genes. 
    
    Inference focuses on interpreting biomarker comparisons between the   disease subtypes IDC and ILC in the context of a larger disease population in the U.S., e.g.,  SEER breast cancer patients \citep{SEER2023}. The estimands of interest include   contrasts and  gene-gene pairwise correlations, alongside disease subtype-specific summaries (e.g., means, standard deviations, and medians).  
    Understanding gene expression and co-expression patterns in different subtypes of breast cancer among national-level patients is crucial for developing feasible guidelines for regulating targeted therapies and  precision medicine \citep{schmidt2016precision}. 
As revealed  in Supplementary Materials, naive group comparisons based on the TCGA patient cohorts are severely confounded by the high degree of covariate imbalance between the IDC and ILC subtypes.

More broadly, covariate balance  is vitally important in observational studies where interest focuses on unconfounded causal comparisons of group potential outcomes  \citep{robins1995semiparametric, Rubin_2007} in a large \textit{natural population} such as the U.S. population. The  \textit{observed populations} of convenience samples such as  observational studies are usually  unrepresentative of this natural population. Theoretical and simulation studies  have demonstrated the conceptual and practical advantages of weighting over other covariate-balancing techniques like matching and regression adjustment \citep{austin2010performance}. As a result,  weighting methods have  widespread applicability in  diverse research areas such as  political science, sociology, and healthcare  \citep{Lunceford_Davidian_2004}. 
For analyzing  cohorts consisting of two groups,  
  the  propensity score  (PS) 
 \citep{Rosenbaum_Rubin_1983} plays a central role. In these studies,  the average treatment effect (ATE) and  average treatment effect on the treated group (ATT) are overwhelmingly popular  estimands  \citep{robins2000marginal}.   
However, the inverse probability weights (IPW) on which these  estimators  rely may be  unstable when some  PSs are near 0 or 1 \citep{li2019propensity}.

Several researchers have proposed variations of  ATE based on truncated
subpopulations of scientific or 
statistical interest \citep{crump2006moving,li2013weighting}. 
{Most weighting methods,  implicitly or explicitly,   provide unbiased inferences  for a specific \textit{pseudo-population}, a covariate-balanced  construct  that often differs substantially from the real but mostly unknown 
 natural population of interest.}
For example, \cite{Li_etal_2018} showed that IPWs correspond to a \textit{combined} pseudo-population and  introduced  the \textit{overlap} pseudo-population,   
wherein the  weights minimize the asymptotic variance
of the weighted average treatment effect for
the overlap pseudo-population (ATO).    For single observational studies comprising two or more groups, 
 \cite{li2019propensity} proposed the \textit{generalized overlap} pseudo-population that minimizes 
the  sum of asymptotic variances of   weighted estimators of  
 pairwise group differences. For multiple observational studies with two  groups, 
  \cite{wang_rosner_2019}   developed an integrative approach for Bayesian  inferences on~ATE. For  single observational studies with two  groups, \cite{mao2019propensity}   obtained analytical variance expressions of modified IPW estimators  adjusted for the estimated  
PS   and augmented the estimators  with outcome models  for improved efficiency. \cite{zeng2023propensity} explored weighting procedures in  single  study,   multiple  group settings with censored survival outcomes.

However,  these methods have several limitations. \textit{First}, 
they are theoretically guaranteed to be effective for a  specific set of   outcome types and estimands under certain theoretical conditions (e.g., equal variances of univariate  group-specific  outcomes). As study endpoints  may be continuous, categorical,   or multivariate, inference procedures for   disparate outcome types have been inadequately explored. Further,
 scientific  interests may necessitate alternative estimands than ATE, ATT or ATO, such as  distribution percentiles, standard deviations,     pairwise correlations of multivariate outcomes, and  unplanned estimands suggested during  post hoc analyses. \textit{Second},
 these methods  may imply  group  
 assignment changes for some subjects that are sometimes difficult to justify for a meaningful, generalizable pseudo-population \citep{Li_etal_2018,li2019propensity}. 
 \textit{Lastly}, very few methods can accommodate the integration of multiple observational studies with multiple unbalanced  groups as encountered in the TCGA datasets. One potential use of the existing weighing methods to achieve covariate balance is by creating a new categorical variable that combines study and group information. However,  it is unclear how to conduct unconfounded group comparisons independent of the ``nuisance" study factor. Furthermore,  the pseudo-populations generated by this approach are often impractical, and inferential accuracies for common estimands are frequently suboptimal.  
There is a critical need for developing efficient  approaches that enable   the integration of multiple observational studies and multiple unbalanced  groups and the construction of pseudo-populations that resemble the natural population of interest.

To fill this gap, we extend  the propensity score  to  the multiple propensity score and propose a new class of  pseudo-populations and multi-study balancing  weights to effectuate data integration and causal meta-analyses. Compared to the existing weighting methods,  our work presents two main advances. First, our framework enables
   unconfounded  inferences on a wide variety of  population-level group features as well as planned or unplanned estimands relevant to  group comparisons.  Second, the framework allows us to derive efficient estimators within this proposed family of  pseudo-populations. Specifically, by maximizing the effective sample size, we further obtain a 
 \underline{FLEX}ible, \underline{O}ptimized, and \underline{R}ealistic (FLEXOR)
   weighting method and derive  new  weighted estimators which  are efficient for a variety of  quantitative, categorical, and multivariate outcomes,   are applicable to different weighting strategies, and effectively utilize   multivariate outcome information. For example, the   estimators 
 yield efficient estimates of  various functionals of  group-specific potential outcomes, e.g., contrasts of means and medians, correlations,  and percentiles.

The rest of the paper is organized as follows.  
 Section~\ref{S:stageI} introduces some basic notation, theoretical assumptions, and a  general covariate-balancing framework for meta-analysis. We further introduce FLEXOR, an optimized pseudo-population, as its special case.  
 Section \ref{S:stageII}  
 develops  unconfounded integrative estimators  applicable to different weighting methods, estimands, and response types, and  establishes asymptotic properties. 
 Section \ref{S:simulation 2} presents the finite sample performance of the proposed methodology, especially when used in conjunction with the FLEXOR weights. Section \ref{S:data analysis}  meta-analyzes the aforementioned  
 TCGA studies and detects differential targeted gene expression and co-expression patterns across the two major breast cancer subtypes in the Unites States.  
 Section \ref{S:discussion} concludes with some final remarks.

%%%%%%%%%%%%%%%%%%%%%
%%%%%%%%%%%%%%%%%%%%%

\section{Integration of  Observational Studies with Multiple Unbalanced  Groups}\label{S:stageI}

\subsection{Notation and basic assumptions}\label{S:notation}

We aim to compare  $K$ subpopulations or groups (e.g., disease subtypes) of participants  belonging to a large natural population such as the U.S.\ patient population.  Beyond  basic summaries (e.g.,  group prevalences) from preexisting registries, no additional information is available about the  natural population. 
The investigation comprises $J$ observational studies.  We assume  $J$ and $K$ are not large.  For $i=1\ldots,N$, let  $Z_i \in \{1, \ldots, K\}$ denote the group and $S_i \in \{1, \ldots, J\}$  denote the 
  observational study. We assume that each participant  belongs to exactly one observational study and each study 
 includes at least one participant in each group. Additionally, there are $p$    covariates  shared by all the studies and  denoted by  $\mathbf{X}_i \in \mathcal{X} \subset \mathcal{R}^p$  for  the $i$th participant.  The motivating TCGA database comprises $K=2$ groups corresponding to breast cancer subtype IDC and ILC, and $p=30$ covariates of    $N=450$  breast cancer patients in $J=7$ observational studies.   
 The $i$th participant's  potential 
outcome is 
$\mathbf{Y}_i^{(z)}=(Y_{i1}^{(z)},\ldots,Y_{iL}^{(z)})' \in \mathcal{R}^L$, i.e.,
the outcome had the patient belonged to group $z=1, \ldots, K.$
The observed outcome  is $\mathbf{Y}_i= \mathbf{Y}_i^{(Z_i)}$.  
In the TCGA example,  vectors $\mathbf{Y}^{(1)}_i$ and $\mathbf{Y}^{(2)}_i$   represent counterfactual  
mRNA measurements of  disease subtypes IDC and ILC on  $L=8$ targeted genes, and  the   observed   $\mathbf{Y}_i \in \mathcal{R}^8$ contains  mRNA measurements of  breast cancer subtype $Z_i$ with which participant $i$ is actually diagnosed.

The participant-specific measurements are a random sample from an   \textit{observed   distribution},  \sg{$p_{+}[S,Z,\mathbf{X},\mathbf{Y}]$}, where  \sg{$p_{+}[\cdot]$} generically represents distributions or densities with respect to the observed population. 
 Extending \cite{Rubin_2007} and \cite{Imbens_2000},
we assume (A)~\textbf{Stable unit treatment value assumption (SUTVA)}:  Given subjects' covariates, the study and group memberships do not influence the potential outcomes\sg{, and no two versions of grouping lead to different potential outcomes}; (B) \textbf{Study-specific  unconfoundedness}: Given  study $S$ and covariate vector  $\mathbf{X}$, group membership $Z$  is independent of  the potential   outcomes $\mathbf{Y}^{(1)}, \ldots, \mathbf{Y}^{(K)}$; and (C) \textbf{Positivity}: Joint density   \sg{$p_{+}[S=z,Z=z,\mathbf{X}=\mathbf{x}]$} is   strictly positive for all $(s,z,\mathbf{x})$. 
 Assumption~(B) states that
  \sg{$p_{+}[\mathbf{Y}^{(z)} \mid S,Z, \mathbf{X}] = p_{+}[\mathbf{Y}^{(z)} \mid S, \mathbf{X}]$}. 
Assumption~(C)  guarantees that the study and  group memberships and   covariates do not have deterministic relationships and often holds  when $J$ and $K$ are not large.

\subsection{A new family of pseudo-populations}\label{S:family} 

We first extend  variations of the propensity score \citep[e.g.,][]{Rosenbaum_Rubin_1983}   to  the  {\textit{multiple propensity score} (MPS)} of the  vector~$(S,Z)$. For  $\mathbf{x} \in \mathcal{X} \subset \mathcal{R}^p$, the MPS
\begin{equation}
    \delta_{sz}(\mathbf{x})    =  
p_{+}\bigl[S=s,Z=z \mid \mathbf{X} = \mathbf{x}\bigr]\quad\text{for   $(s,z) \in \Sigma$
$\equiv$ $\{1,\ldots,J\} \times \{1,\ldots,K\}.$
} \label{o-MPS}
\end{equation}
It then follows that the   joint  density   \sg{$p_{+}\bigl[S=s,Z=z,\mathbf{X}=\mathbf{x}\bigr]=\delta_{sz}(\mathbf{x}) f_+(\mathbf{x})$}, where $f_+(\mathbf{x})=$ \sg{$p_{+}[\mathbf{X}=\mathbf{x}]$} represents the marginal  covariate density in the observed population. As the MPS is unknown in observational studies, we can estimate it by regressing  the combinations of $(S_i,Z_i)$ on  covariate $\mathbf{x}_i$  ($i=1,\ldots,N$). 
 In single studies with two groups, the PS is usually estimated using logistic regression \citep{mao2019propensity}. For estimating  MPS, we recommend    multinomial logistic regression: $\log \bigl(\delta_{sz}(\mathbf{x})/\delta_{11}(\mathbf{x})\bigr)=\boldsymbol{\omega}_{sz}'\mathbf{x}$ for $(s,z) \neq (1,1)$, so that  $\boldsymbol{\omega}=\bigl\{\boldsymbol{\omega}_{sz}:(s,z) \neq (1,1)\bigr\}$ is a $(JK-1)p$-dimensional  parameter. If we define $\boldsymbol{\omega}_{11}$ to be the vector of $p$ zeros, then $\delta_{sz}(\mathbf{x})=\exp(\boldsymbol{\omega}_{sz}'\mathbf{x})/\sum_{s^*=1}^J\sum_{z^*=1}^K\exp(\boldsymbol{\omega}_{s^*z^*}'\mathbf{x})$ for all  $(s,z) \in \Sigma$.

Consider  a pseudo-population with    attributes fully or partially 
 prescribed by the investigator via two  probability vectors:  (i) 
     relative amounts of information  extracted from the  studies, quantified by  probability tuple $\boldsymbol{\gamma}=(\gamma_1,\ldots,\gamma_J)$; 
     and (ii) 
       relative group prevalence, $\boldsymbol{\theta}=(\theta_{1},\ldots,\theta_{K})$.  
For instance, in the  TCGA breast cancer studies, setting  $\gamma_j=1/7$  extracts equal  information from each  study, whereas  $\boldsymbol{\theta}=$ $(8/9,  1/9)$ constrains the pseudo-population to the known U.S. proportions  of  breast cancer subtypes IDC and ILC      \citep{IDC,ILC}.  
 If  some or all components of $\boldsymbol{\gamma}$ or $\boldsymbol{\theta}$ are unknown, 
   subsequent inferences  can optimize the pseudo-population over the multiple possibilities for these quantities.

For multiple observational studies, the participant study memberships $S_1,\ldots,S_N$  are primarily influenced by the $J$ study designs and  unknown factors  driving  participation;  moreover,   study participant characteristics can differ substantially across studies, 
especially in cancer investigations. 
To address  these issues, we aim to design a  pseudo-population for achieving  theoretical covariate balance between the $K$ groups.  In other words,  we construct a  pseudo-population wherein the study memberships, the group memberships and   patient characteristics are mutually independent,  
i.e., $S \perp Z \perp \mathbf{X},$ so that 
\begin{equation}
   p\bigl[S=s, Z=z,\mathbf{X}=\mathbf{x}\bigr]= \gamma_s \, \theta_z \,  f_{\boldsymbol{\gamma},\boldsymbol{\theta}}(\mathbf{x}),  \quad \text{for $(s,z,\mathbf{x}) \in \Sigma\times\mathcal{X}$.}\label{pseudo1}
\end{equation}
 Here and hereafter, \sg{$p[\cdot]$} denotes a distribution or  density  with respect to the designed pseudo-population, whereas \sg{$p_{+}[\cdot]$} corresponds to the observed population, as mentioned earlier. Equation~(\ref{pseudo1}) further emphasizes that although $S$, $Z$, and $\mathbf{X}$ are independent  in the pseudo-population, they may share some distributional parameters. More explicitly, the subscripts of $f_{\boldsymbol{\gamma},\boldsymbol{\theta}}(\mathbf{x})$ emphasize that the  pseudo-population  density of  $\mathbf{X}$ may depend on  $\boldsymbol{\gamma}$ and $\boldsymbol{\theta}$.

 Next,   consider the relationship between the  pseudo-population covariate density, $f_{\boldsymbol{\gamma},\boldsymbol{\theta}}(\mathbf{x})$, and the marginal observed covariate density, $f_{+}(\mathbf{x})$. Assuming a common dominating measure for the densities and a common support, $\mathcal{X}$,  there exists without loss of generality a positive \textit{tilting function} \citep[e.g.,][]{Li_etal_2018} denoted by $\eta_{\boldsymbol{\gamma},\boldsymbol{\theta}}$   such that
$f_{\boldsymbol{\gamma},\boldsymbol{\theta}}(\mathbf{x}) \propto \eta_{\boldsymbol{\gamma},\boldsymbol{\theta}}(\mathbf{x})f_{+}(\mathbf{x})$ for all $\mathbf{x} \in \mathcal{X}$. Therefore, 
  $f_{\boldsymbol{\gamma},\boldsymbol{\theta}}(\mathbf{x}) = \eta_{\boldsymbol{\gamma},\boldsymbol{\theta}}(\mathbf{x})f_{+}(\mathbf{x})/\mathbb{E}_+[\eta_{\boldsymbol{\gamma},\boldsymbol{\theta}}(\mathbf{X})]$ where  $\mathbf{X} \sim f_{+}$ and $\mathbb{E}_+(\cdot)$ denotes expectations under the observed distribution. Intuitively,  high tilting function values correspond to covariate space regions  with   high  pseudo-population~weights. Let $\mathcal{S}_{J}$ denote the unit simplex in $\mathcal{R}^J$. Different choices of    $\boldsymbol{\gamma}\in\mathcal{S}_{J}$, $\boldsymbol{\theta}\in\mathcal{S}_{K}$,  and  tilting function $\eta_{\boldsymbol{\gamma},\boldsymbol{\theta}}$     identify different  pseudo-populations with structure  (\ref{pseudo1}). 

\paragraph{Balancing  weights for integration of multiple studies}  To efficiently meta-analyze 
multiple studies (with $J >1$), we propose the  \textit{multi-study balancing  weight}, defined  
as the ratio of the joint densities with respect to the pseudo-population and  observed population. More specifically, for any $(s,z,\mathbf{x}) \in \Sigma\times\mathcal{X}$, the multi-study balancing  weight
\begin{equation}
    \rho_{\boldsymbol{\gamma},\boldsymbol{\theta}}(s,z,\mathbf{x}) = \frac{p\bigl[S=s,Z=z,\mathbf{X}=\mathbf{x}\bigr]\hfill}{ p_{+}\bigl[S=s,Z=z,\mathbf{X}=\mathbf{x}\bigr]} 
    = 
\frac{\gamma_s \, \theta_z  \, f_{\boldsymbol{\gamma},\boldsymbol{\theta}}(\mathbf{x})}{\delta_{sz}(\mathbf{x}) \,f_{+}(\mathbf{x})} = 
\frac{\gamma_s \, \theta_z  \, \eta_{\boldsymbol{\gamma},\boldsymbol{\theta}}(\mathbf{x})}{\delta_{sz}(\mathbf{x}) \,\mathbb{E}_+[\eta_{\boldsymbol{\gamma},\boldsymbol{\theta}}(\mathbf{X})]}. 
\label{w1}
\end{equation}
As $\rho_{\boldsymbol{\gamma},\boldsymbol{\theta}}(s,z,\mathbf{x}) \times p_{+}\bigl[S=s,Z=z,\mathbf{X}=\mathbf{x}\bigr]
= p\bigl[S=s,Z=z,\mathbf{X}=\mathbf{x}\bigr]$, 
the balancing  weight serves to redistribute the   observed distribution's relative mass to match  that of the pseudo-population.  Defining the \textit{unnormalized weight function} as   $\tilde{\rho}_{\boldsymbol{\gamma},\boldsymbol{\theta}}(s,z,\mathbf{x})=$ $\gamma_s  \theta_z   \eta_{\boldsymbol{\gamma},\boldsymbol{\theta}}(\mathbf{x})/\delta_{sz}(\mathbf{x})$,  the  \textit{unnormalized      weight} of the $i$th participant is  $\tilde{\rho}_i=$ $\tilde{\rho}_{\boldsymbol{\gamma},\boldsymbol{\theta}}(s_i,z_i,\mathbf{x}_i)$. 
For a  general pseudo-population (e.g.,  FLEXOR pseudo-population  introduced in the sequel), the unnormalized weights, even  within a study-group combination, may depend on $\boldsymbol{\gamma}$ and $\boldsymbol{\theta}$ through the tilting function. As discussed  later, the unnormalized    weights can be utilized to provide  unconfounded  inferences on various potential outcome features for a general pseudo-population. 

The proposed pseudo-populations and balancing  weights are general, encompassing many well-known weighting methods  in  single-study settings.  For example, in single studies, assume    equally prevalent pseudo-population groups ($\theta_z=1/K$) in   expression (\ref{pseudo1}). A constant tilting function 
 yields
   IPWs when $K=2$ and generalized IPWs \citep{Imbens_2000} when $K>2$. On the other hand,   $\eta_{\boldsymbol{\gamma},\boldsymbol{\theta}}(\mathbf{x}) =1/\sum_{z}\delta_z^{-1}(\mathbf{x})$    produces  overlap weights    \citep{Li_etal_2018} when $K=2$, and     generalized overlap weights \citep{li2019propensity} when $K>2$. Again,  if $\eta_{\boldsymbol{\gamma},\boldsymbol{\theta}}(\mathbf{x}) = \delta_{z'}(\mathbf{x})$ for a  group $z'$, then the pseudo-population's covariate density, $f_{\boldsymbol{\gamma},\boldsymbol{\theta}}(\mathbf{x})$, matches the  observed covariate density of  the    group $z'$  participants.

The choice of  different tilting functions in  (\ref{pseudo1}) naturally extends   several  weighting methods designed for single studies to meta-analytical settings.  For example,  assuming equally weighted studies and equally prevalent groups, i.e., $\gamma_s=1/J$ and $\theta_z=1/K$, a constant tilting function $\eta_{\boldsymbol{\gamma},\boldsymbol{\theta}}(\mathbf{x})\propto 1$ and  $\eta_{\boldsymbol{\gamma},\boldsymbol{\theta}}(\mathbf{x}) =1/\sum_s\sum_z\delta_{sz}^{-1}(\mathbf{x})$, respectively,  produces    extensions of the combined \citep{Li_etal_2018} and generalized overlap \citep{li2019propensity} pseudo-populations  
 appropriate for  meta-analyzing multiple studies with multiple groups. We refer to these proposed pseudo-populations as the \textit{integrative combined} (IC) 
and \textit{integrative generalized overlap} (IGO) pseudo-populations, respectively. Similarly, for a fixed group $z'$,  the tilting function $\eta_{\boldsymbol{\gamma},\boldsymbol{\theta}}(\mathbf{x}) =\sum_{s}\delta_{sz'}(\mathbf{x})$ gives a  pseudo-population whose marginal covariate density equals the  observed covariate density of  group $z'$  participants irrespective of their study memberships.   Given the availability of different tilting functions, an important question arises: which choice is  optimal  and in what sense? We address this below.

\paragraph{Effective sample size} A widely used measure 
of a pseudo-population's  inferential accuracy  
is the \textit{effective sample size} (ESS), 
$  \mathcal{Q}(\boldsymbol{\gamma}, \boldsymbol{\theta},\eta_{\boldsymbol{\gamma},\boldsymbol{\theta}}) = N/\bigl[1+\text{Var}_{+}\bigl\{\rho_{\boldsymbol{\gamma},\boldsymbol{\theta}}(S,Z,\mathbf{X})\bigr\}\bigr] 
    = N/\mathbb{E}_{+}\bigl\{\rho_{\boldsymbol{\gamma},\boldsymbol{\theta}}^2(S,Z,\mathbf{X})\bigr\}, 
    $
which relies on the second moment (provided it exists) of the  balancing weights in the observed population \citep[e.g.,][]{mccaffrey2013tutorial}.  The ESS is asymptotically equivalent to the sample ESS,  $\tilde{\mathcal{Q}}(\boldsymbol{\gamma}, \boldsymbol{\theta},\eta_{\boldsymbol{\gamma},\boldsymbol{\theta}}) = N^2(\sum_{i=1}^N \tilde{\rho}_{i})^2/\sum_{i=1}^N \tilde{\rho}_{i}^2$.  Informally, the ESS is the hypothetical sample size from the  pseudo-population  containing the same information as $N$  samples from the  observed population, and it is always less than $N$ unless the pseudo-population and observed population are identical.

%%%%%%%%%%%%%%%%%%%%%%%%%%%%%%%%%%%%

\paragraph{An optimized case: FLEXOR pseudo-population} 
 We propose 
 \textit{FLEXOR} as a member of  pseudo-population family (\ref{pseudo1})  that maximizes the ESS or minimizes the variation of the balancing weights, subject to any problem-dictated constraints on the vectors $\boldsymbol{\gamma}$ and $\boldsymbol{\theta}$. That is, if  the  triplet $(\breve{\boldsymbol{\gamma}}, \breve{\boldsymbol{\theta}},\breve{\eta}_{\breve{\boldsymbol{\gamma}}, \breve{\boldsymbol{\theta}}})$ identifies the FLEXOR pseudo-population and   
$(\boldsymbol{\gamma}, \boldsymbol{\theta})$ is known to belong to a subset, $\Upsilon$, of $\mathcal{S}_J \times \mathcal{S}_K$, then 
$\mathcal{Q}(\breve{\boldsymbol{\gamma}}, \breve{\boldsymbol{\theta}},\breve{\eta}_{\breve{\boldsymbol{\gamma}}, \breve{\boldsymbol{\theta}}}) = \sup_{(\boldsymbol{\gamma}, \boldsymbol{\theta}) \in \Upsilon} \sup_{\eta_{\boldsymbol{\gamma},\boldsymbol{\theta}}} \mathcal{Q}(\boldsymbol{\gamma}, \boldsymbol{\theta},\eta_{\boldsymbol{\gamma},\boldsymbol{\theta}})$.

\paragraph{A two-step procedure for constructing the FLEXOR pseudo-population}  
 Starting with an initial  $(\boldsymbol{\gamma}, \boldsymbol{\theta})\in\Upsilon$,  we  
 iteratively perform the following   steps until convergence:

\begin{itemize}
    \item \textbf{Step I} \quad  For a fixed  $(\boldsymbol{\gamma}, \boldsymbol{\theta})$, maximize  {sample ESS $\tilde{\mathcal{Q}}(\boldsymbol{\gamma}, \boldsymbol{\theta},\eta_{\boldsymbol{\gamma},\boldsymbol{\theta}})$} over all  tilting functions, $\eta_{\boldsymbol{\gamma},\boldsymbol{\theta}}$. This gives the \textit{best fixed-$(\boldsymbol{\gamma}, \boldsymbol{\theta})$ pseudo-population} identified by   $(\boldsymbol{\gamma}, \boldsymbol{\theta},\breve{\eta}_{\boldsymbol{\gamma},\boldsymbol{\theta}})$. The analytical form of $\breve{\eta}_{\boldsymbol{\gamma},\boldsymbol{\theta}}$  {for the theoretical ESS} is given in Theorem \ref{Thm1} below. Set function $\eta=\breve{\eta}_{\boldsymbol{\gamma},\boldsymbol{\theta}}$.
  
  \item  \textbf{Step II} \quad  For a fixed  tilting function $\eta$, maximize  {$\tilde{\mathcal{Q}}(\boldsymbol{\gamma}, \boldsymbol{\theta},\eta)$} over all $(\boldsymbol{\gamma}, \boldsymbol{\theta})\in\Upsilon$ to obtain  the  \textit{best fixed-$\eta$ pseudo-population}, identified by the triplet $(\tilde{\boldsymbol{\gamma}}, \tilde{\boldsymbol{\theta}},\eta)$. This  parametric maximization over $\Upsilon\subset \mathcal{S}_J \times \mathcal{S}_K$ can be quickly performed in R using the {\tt optim} function or  by  Gauss-Seidel or  Jacobi  algorithms.  Set $(\boldsymbol{\gamma}, \boldsymbol{\theta})=(\tilde{\boldsymbol{\gamma}}, \tilde{\boldsymbol{\theta}})$.
 
\end{itemize}

In our experience,  convergence is attained within only a few iterations.  
 The converged pseudo-population  with the largest ESS  yields the FLEXOR pseudo-population. The following theorem gives the analytical expression for the global maximum of $\mathcal{Q}(\boldsymbol{\gamma}, \boldsymbol{\theta},\eta_{\boldsymbol{\gamma},\boldsymbol{\theta}})$ mentioned in Step~I. See  Supplementary Materials  for the proof.

 \begin{theorem}\label{Thm1}  Suppose   probability vectors $\boldsymbol{\gamma}$ and $\boldsymbol{\theta}$ have  
  strictly positive elements  and are held fixed. Let $\boldsymbol{\Xi}$ be the set of tilting functions    for which the ESS, $\mathcal{Q}(\boldsymbol{\gamma}, \boldsymbol{\theta},\eta_{\boldsymbol{\gamma},\boldsymbol{\theta}})$, of  pseudo-population  (\ref{pseudo1}) is finite.  Maximizing    $\mathcal{Q}(\boldsymbol{\gamma}, \boldsymbol{\theta},\eta_{\boldsymbol{\gamma},\boldsymbol{\theta}})$ over  all  tilting functions $\eta_{\boldsymbol{\gamma},\boldsymbol{\theta}}\in\boldsymbol{\Xi}$, the optimal fixed-$(\boldsymbol{\gamma}, \boldsymbol{\theta})$ pseudo-population's tilting function,   denoted by $\breve{\eta}_{\boldsymbol{\gamma},\boldsymbol{\theta}}$, has the expression:
  \begin{equation}
       \breve{\eta}_{\boldsymbol{\gamma},\boldsymbol{\theta}}(\mathbf{x}) =   \biggl( \sum_{s=1}^J  \sum_{z=1}^K \frac{\gamma_{s} ^2\theta_z^2}{\delta_{sz}(\mathbf{x})}  \biggr)^{-1},  \quad \text{ $\mathbf{x} \in \mathcal{X}$.}\label{optimal_psi}
  \end{equation}
  The unnormalized  weight function  for the optimal fixed-$(\boldsymbol{\gamma}, \boldsymbol{\theta})$ pseudo-population is then
  \begin{equation}
 \tilde{\rho}_{\boldsymbol{\gamma},\boldsymbol{\theta}}(s,z,\mathbf{x}) 
= \frac{1}{\gamma_s\theta_z}  
\biggl(\frac{\gamma_s^2\theta_z^2/\delta_{sz}(\mathbf{x}) }{ \sum_{t=1}^J \sum_{u=1}^K \gamma_t^2\theta_u^2/\delta_{tu}(\mathbf{x})}\biggr),\qquad\text{for $(s,z) \in \Sigma$ and $\mathbf{x} \in \mathcal{X}$.}
\end{equation}
The optimal fixed-$(\boldsymbol{\gamma}, \boldsymbol{\theta})$ pseudo-population's balancing  weights, evaluated as in (\ref{w1}),     are uniformly bounded. 
The ESS of the  optimal fixed-$(\boldsymbol{\gamma}, \boldsymbol{\theta})$ pseudo-population is $N \mathbb{E}_{+}\bigl[ \breve{\eta}_{\boldsymbol{\gamma},\boldsymbol{\theta}}(\mathbf{X})\bigr] $ with the expectation taken   over  $\mathbf{X} \sim f_{+}(\mathbf{x})$,  the observed population's covariate density.  
 
 \end{theorem}

It can be shown that the optimal tilting function $\breve{\eta}_{\boldsymbol{\gamma},\boldsymbol{\theta}}(\mathbf{x})$ apportions  low importance to outlying regions of  covariate space $\mathcal{X}$ where $\delta_{sz}(\mathbf{x})$ is approximately 0 for some $(s,z) \in \Sigma$. Furthermore, the optimal tilting function   emphasizes  covariate regions where the group propensities $\delta_{1}(\mathbf{x}),\ldots,\delta_{K}(\mathbf{x})$ match the group proportions $\theta_1,\ldots,\theta_K$ of the  larger natural population. In particular, in  pseudo-populations with equally prevalent groups,  the  tilting function    promotes covariate regions where the group propensities are approximately equal.
%%%%%%%%%%%%%%%%%%%%%%%%%%%%%%%%%%%%
 %%%%%%%%%%%%%%%%%%%%%%%%%%%%%%%%%%%%%%%%%%%%%%%%%%

\section{Meta-analyses of  Group Potential Outcomes}\label{S:stageII}

Causal meta-analyses  
generally follow a two-stage inferential procedure \citep[e.g.,][]{Rubin_2007}. In \textit{Stage 1}, the ``outcome free'' analysis only  utilizes   covariate information to   estimate  the  propensity scores, as done  in Section~\ref{S:stageI}. In \textit{Stage 2}, for the   pseudo-population of interest, the procedure makes unconfounded  comparisons of   group potential outcomes via estimands such as pairwise difference of group means. For any known pseudo-population  belonging to  family (\ref{pseudo1}), 
 the procedure  accommodates  wide-ranging group-level features of the endpoints using the available multivariate outcome information. Additionally,  we derive analytical expressions for the asymptotic variances of the proposed multivariate estimators.

   Suppose   
  potential outcome vectors $\mathbf{Y}^{(1)},\ldots,$ $\mathbf{Y}^{(K)}$ have a common support, $\mathcal{Y}\subset\mathcal{R}^L$. To ensure  that  SUTVA,  unconfoundedness, and positivity   of the observed population  also  hold for  the pseudo-population, we  assume  identical conditional distributions: 
 \begin{equation}
     \sg{p[\mathbf{Y}^{(z)} \mid S,Z, \mathbf{X}] = p_{+}[\mathbf{Y}^{(z)} \mid S,Z,  \mathbf{X}]} \quad\text{for group $z=1,\ldots,K$}, \label{Y_z|S,X}
 \end{equation}
 where  \sg{$p_{+}[\cdot| \cdot]$ and $p[\cdot| \cdot]$}  denote the  observed and pseudo-population conditional densities, respectively.  Unlike the observed population, the covariate-balanced  pseudo-population
entails \sg{$
   p [\mathbf{Y} \mid Z=z] = p[\mathbf{Y}^{(z)}]$},
 enabling us  to construct  weighted estimators of various features of the pseudo-population potential outcomes.

 Let $\mathbb{E}[\cdot]$   denote expectations with respect to the  pseudo-population. Let $\Phi_1,\ldots,\Phi_M$  be real-valued 
 functions having domain $\mathcal{Y}$. We wish to infer 
 pseudo-population means of   transformed potential  outcomes, $\mathbb{E}[\Phi_1(\mathbf{Y}^{(z)})],\ldots,\mathbb{E}[\Phi_M(\mathbf{Y}^{(z)})]$  for   $z=1,\ldots, K$.  Appropriate choices of   $\Phi_m$ correspond to pseudo-population inferences about group-specific  marginal
means, medians,  variances, and  CDFs of     potential outcome components. 
Equivalently, writing $\mathbf{\Phi}(\mathbf{Y}^{(z)})=\bigl(\Phi_1(\mathbf{Y}^{(z)}),\ldots,\Phi_M(\mathbf{Y}^{(z)})\bigr)' \in \mathcal{R}^M$, the  inferential focus is the vector, $\boldsymbol{\lambda} ^{(z)}=$ $\mathbb{E}[\mathbf{\Phi}(\mathbf{Y}^{(z)})]$. 

 For  real-valued functions $\psi$ with domain $\mathcal{R}^M$, 
we estimate $\psi(\boldsymbol{\lambda} ^{(z)})$. 
For example, if the   first two components of $\mathbf{Y}^{(z)}$  are quantitative, then defining  $\Phi_1(\mathbf{Y}^{(z)})=Y_1^{(z)}$, $\Phi_2(\mathbf{Y}^{(z)})=Y_2^{(z)}$,  $\Phi_3(\mathbf{Y}^{(z)})=Y_1^{(z)}Y_2^{(z)}$, and  $\psi(t_1,t_2,t_3)=t_3-t_1t_2$, we obtain $\psi(\boldsymbol{\lambda} ^{(z)})=\text{cov}(Y_1^{(z)},Y_2^{(z)})$ as the pseudo-population covariance of $Y_1^{(z)}$ and $Y_2^{(z)}$ in the  $z$th group. The pseudo-population correlation  of  pairwise  components of $\mathbf{Y}^{(z)}$ can be estimated from estimates of the covariance and standard deviations, as in the motivating breast cancer studies, where the goal is to estimate the pairwise correlations of the eight targeted genes in groups $z\in\{1,2\}$ (i.e., IDC and ILC  subtypes). For a second example, let $y_{11},\ldots,y_{1M}$ be  a fine grid of prespecified points in the support of the first component $Y_1^{(z)}$  and  $\Phi_m(\mathbf{Y}^{(z)})=\mathcal{I}(Y_1^{(z)}\le y_{1m})$. For  $\psi(t_1,\ldots,t_M)=t_{m}$,   the pseudo-population CDF of $Y_1^{(z)}$ evaluated at $y_{1m}$ equals  $\psi(\boldsymbol{\lambda} ^{(z)})$. Similarly, for  $\psi(t_1,\ldots,t_M)=t_{m^*}$ where $m^*=\arg\min_m |t_m-0.5| $,  the approximate pseudo-population median of $Y_1^{(z)}$ is given by $\psi(\boldsymbol{\lambda} ^{(z)})$.

Using the unnormalized   weights $\tilde{\rho}_1,\ldots, \tilde{\rho}_N$
[defined underneath equation (\ref{w1})] of a pseudo-population, 
we estimate  $\mathbb{E}[\mathbf{\Phi}(\mathbf{Y}^{(z)})]$ as  random vector 
\begin{equation}
\mathbf{\bar{\Phi}}_{z} = \frac{\sum_{i=1}^N \tilde{\rho}_i\,\mathbf{\Phi}(\mathbf{Y}_i)\,\mathcal{I}(Z_i=z)}{\sum_{i=1}^N \tilde{\rho}_i\,\mathcal{I}(Z_i=z)}. \label{Yz}
\end{equation}
The following theorem and corollaries study  asymptotic properties of  random vector $\mathbf{\bar{\Phi}}_{z}$ as an estimator of  multivariate feature  $\mathbb{E}[\mathbf{\Phi}(\mathbf{Y}^{(z)})]$. Part \ref{Thm:estimation,part 2a} of the theorem 
considers a simpler  situation where the MPS is known. As discussed in  \cite{mao2019propensity} and \cite{zeng2023propensity},  Part \ref{Thm:estimation,part 2b}   considers a more  realistic situation in which the MPS is estimated. 
The proofs   are available in Supplementary Materials.

 \begin{theorem}\label{Thm:estimation}
 Let $\mathbb{E}_+[\cdot]$ and $\mathbb{E}[\cdot]$  respectively denote expectations with respect to the observed population and a pseudo-population of the form (\ref{pseudo1}).
  Let  observed   probability $P_+[S=s]$ be strictly positive for study $s=1,\ldots,J$.
Suppose the conditional distributions of the potential outcomes are  unconfounded, as described in Section \ref{S:stageI}, and satisfy assumption~(\ref{Y_z|S,X}).  Suppose the multi-study balancing  weight (\ref{w1}) is such that $\mathbb{E}_+[\rho_{\boldsymbol{\gamma},\boldsymbol{\theta}}^2(S,Z,\mathbf{X})]$   is finite. For  $m=1,\ldots,M$, let
     $\Phi_m$ be a  real-valued function with domain $\mathcal{Y}$ such that  $\mathbb{E}_+[\rho^2_{\boldsymbol{\gamma},\boldsymbol{\theta}}(S,Z,\mathbf{X})\Phi_m(\mathbf{Y})]$ is finite. 
     For group $z=1,\ldots,K$, interest focuses on  the pseudo-population moment, $\mathbb{E}[\mathbf{\Phi}(\mathbf{Y}^{(z)})]$, also denoted by vector $\boldsymbol{\lambda}^{(z)}=(\lambda_{1}^{(z)},\ldots,\lambda_{M}^{(z)})'$. For estimator $\mathbf{\bar{\Phi}}_{z}$ defined in (\ref{Yz}), as $N \to \infty$:

  \begin{enumerate}[label=(\arabic*), ref=\arabic*]
      \item\label{Thm:estimation,part 1} \textbf{Consistency}:   $\mathbf{\bar{\Phi}}_{z} \stackrel{p}\to \boldsymbol{\lambda}^{(z)} $. 
      
      \item\label{Thm:estimation,part 2} \textbf{Asymptotic normality}: 
     Consider the following situations:
 
 \begin{enumerate}
     \item\label{Thm:estimation,part 2a} \textbf{Known MPS}: \quad Suppose multiple propensity score (\ref{o-MPS}) is known. Let    variance matrix
  \begin{equation*}
      \boldsymbol{\Sigma}_1^{(z)}=\frac{1}{\theta_z^2} \mathbb{E}_{+}\biggl(\rho_{\boldsymbol{\gamma},\boldsymbol{\theta}}^2(S,Z,\mathbf{X})\,\mathcal{I}(Z=z)\,\bigl(\mathbf{\Phi}(\mathbf{Y}_i)-\boldsymbol{\lambda}^{(z)}\bigr)\bigl(\mathbf{\Phi}(\mathbf{Y}_i)-\boldsymbol{\lambda}^{(z)}\bigr)'\biggr).\label{Sigma_z}
  \end{equation*}
     Then $\sqrt{N}\bigl(\mathbf{\bar{\Phi}}_{z} - \boldsymbol{\lambda}^{(z)}\bigr) \stackrel{d}\to N_M\bigl(\mathbf{0}, \boldsymbol{\Sigma}_1^{(z)}\bigr)$.

\smallskip

 \item\label{Thm:estimation,part 2b} \textbf{Estimated MPS}: \quad Suppose   the  MPS is  estimated using multinomial logistic regression as outlined after definition (\ref{o-MPS}). Let $\hat{\boldsymbol{\omega}}$ be the MLE of parameter $\boldsymbol{\omega}$ that determines the unnormalized   weights $\tilde{\rho}_1,\ldots, \tilde{\rho}_N$ in estimator (\ref{Yz}).   We denote the 
 variance matrix   of Part~\ref{Thm:estimation,part 2a}  
  by
 $\boldsymbol{\Sigma}_1^{(z)}(\boldsymbol{\omega})$ to make explicit its  dependence   on $\boldsymbol{\omega}$.
       Define variance matrix 
       $\boldsymbol{\Sigma}_2^{(z)}(\boldsymbol{\omega})=\boldsymbol{\Sigma}_1^{(z)}(\boldsymbol{\omega}) + \mathbf{D}^{(z)}(\boldsymbol{\omega})$, 
       where  $\mathbf{D}^{(z)}(\boldsymbol{\omega})$ is   given in  Supplementary Materials.       Then
   $\sqrt{N}\bigl(\mathbf{\bar{\Phi}}_{z} - \boldsymbol{\lambda}^{(z)}\bigr) \stackrel{d}\to N_M\bigl(\mathbf{0}, \boldsymbol{\Sigma}_2^{(z)}(\boldsymbol{\omega})\bigr)$.

 \end{enumerate}

  \end{enumerate}
 \end{theorem}

\begin{corollary}\label{corollary 2}  
 Suppose   the  MPS is  estimated using multinomial logistic regression. Let  $\psi$ be a real-valued  differentiable function with domain $\mathcal{R}^M$. Let $\nabla \psi(\boldsymbol{\lambda})=\partial \psi(\boldsymbol{\lambda})/\partial \boldsymbol{\lambda}$ denote the gradient vector of length $M$ at $\boldsymbol{\lambda}$. With $\boldsymbol{\lambda}^{(z)}=\mathbb{E}[\mathbf{\Phi}(\mathbf{Y}^{(z)})]$, suppose   gradient vector $\nabla \psi(\boldsymbol{\lambda}^{(z)})$ is non-zero at $\boldsymbol{\lambda}^{(z)}$. With  variance matrix  $\boldsymbol{\Sigma}_2^{(z)}(\boldsymbol{\omega})$  defined as in Part~\ref{Thm:estimation,part 2b} of Theorem~\ref{Thm:estimation}, set $\tau_z(\boldsymbol{\omega})=$ $\nabla' \psi(\boldsymbol{\lambda}^{(z)}) \, \boldsymbol{\Sigma}_2^{(z)}(\boldsymbol{\omega}) \,\nabla \psi(\boldsymbol{\lambda}^{(z)})$. Then $\psi(\mathbf{\bar{\Phi}}_{z})$ is an  asymptotically normal estimator of $\psi(\boldsymbol{\lambda}^{(z)})$:
$\sqrt{N}\bigl(\psi(\mathbf{\bar{\Phi}}_{z}) - \psi(\boldsymbol{\lambda}^{(z)})\bigr) \stackrel{d}\to N\bigl(0, \tau_z^2(\boldsymbol{\omega})\bigr).
$
\end{corollary}

\noindent  
\paragraph{\bf Remark} Theorem \ref{Thm:estimation} and its corollaries summarize several  noteworthy features of estimator (\ref{Yz}), in that it:   (i)  is applicable to the  balancing weights of any pseudo-populations, including  IC, IGO,  and FLEXOR weights; (ii)   generalizes 
plug-in sample moment estimators \citep{li2019propensity} to multiple  groups and  studies, while accommodating mixed-type multivariate outcomes, 
 (iii)   exploits known or researcher-supplied information about the group proportions  of the  pseudo-population; as  mentioned, the FLEXOR weights  typically set  $\theta_z$  equal  to  the  known group prevalences of the larger population. By contrast,  $\theta_z=1/K$ for    most other weighting methods, and (iv)  extends  \cite{mao2019propensity} by quantifying the sampling errors in multiple group settings; matrix $\mathbf{D}^{(z)}(\boldsymbol{\omega})$ in Part \ref{Thm:estimation,part 2b} represents the adjustment due to MPS estimation, and in the event that parameter $\boldsymbol{\omega}$ is known, this adjustment term  vanishes and matrix $\boldsymbol{\Sigma}_2^{(z)}(\boldsymbol{\omega})$ of  Part \ref{Thm:estimation,part 2b} equals $\boldsymbol{\Sigma}_1^{(z)}$ of  Part \ref{Thm:estimation,part 2a}.

\paragraph{Group comparisons} 
  Consider estimation of the pseudo-population moment $\mathbb{E}[\Phi_m(\mathbf{Y}^{(z)})]$ using $\bar{\Phi}_{zm}$.
  Applying standard results \citep[e.g.,][Chapter 5]{johnson2002applied},  we can construct approximate   $100(1-\alpha)$\% confidence intervals \textit{simultaneously} for all possible linear combinations of $\mathbb{E}[\Phi_m(\mathbf{Y}^{(1)})],\ldots,$ $\mathbb{E}[\Phi_m(\mathbf{Y}^{(K)})]$. In particular, for large $N$,  using the estimated $m$th diagonal element, $s^{(z)}_{mm}$, of the variance matrix  defined in Theorem \ref{Thm:estimation},   the interval   $\sum_{z=1}^K a_z\,\bar{\Phi}_{zm} \pm  \sqrt{\chi^2_K(\alpha)\sum_{z=1}^K a_z^2\,s_{mm}^{(z)}/N}$   contains $\sum_{z=1}^K a_z\,\mathbb{E}[\Phi_m(\mathbf{Y}^{(z)})]$ with approximate probability $(1-\alpha)$ simultaneously for all  scalars $a_1,\ldots,a_K$.  
 Various  pseudo-population features can then be  compared between the $K$  groups.
  Writing  $\lambda^{(zm)}=\mathbb{E}[Y_m^{(z)}]$, we could estimate  $\lambda^{(1m)} - \lambda^{(2m)}$ (e.g., {average difference between the $m$th gene's mRNA expression levels  for IDC  and ILC breast cancer patients}) and, when $K>2$, $\lambda^{(1m)} - \frac{1}{K-1}\sum_{z=2}^K \lambda^{(zm)}$ (e.g., {for the $m$th gene, average difference between  the mRNA expression levels for  a reference group  and the average of the other groups}). We could also estimate {ratios of means}  such as $\lambda^{(zm)}/\lambda^{(1m)}$,
{ratios of mean differences} such as $\bigl(\lambda^{(3m)} -\lambda^{(1m)}\bigr)/\bigl(\lambda^{(2m)} -\lambda^{(1m)}\bigr)$,  
{group-specific standard deviations, percentiles,  ratios of  medians, and ratios of coefficients of variation.}  
 Under mild conditions, these estimators are  consistent and asymptotically normal, and their asymptotic variances are available by applying Corollary~\ref{corollary 2}  and  the delta method. If   $N_z$ is small for some groups, such as  rare or undersampled treatments,  the asymptotic  confidence intervals  may not have proper  coverage and 
we could  employ bootstrap methods 
to  construct confidence intervals.

In single studies ($J=1$), \cite{hirano2003efficient} and \cite{zeng2023propensity} have shown that treating IPWs as known  counter-intuitively overestimates the variance of pairwise group mean comparisons. However, with multiple studies and arbitrary functions of group-specific features $\psi(\boldsymbol{\lambda}^{(1)}),\ldots,\psi(\boldsymbol{\lambda}^{(K)})$, this is not generally guaranteed  because  matrix  $\mathbf{D}^{(z)}(\boldsymbol{\omega})$ of Theorem~\ref{Thm:estimation} may be neither positive nor negative definitive for a general pseudo-population~(\ref{pseudo1}). 

%%%%%%%%%%%%%%%%%%%%%%%%%%%%%%

\section{Simulation Study} \label{S:simulation 2}

We used simulated datasets to evaluate  different  weighting strategies  for inferring the    population-level features of two subject groups and assessed the accuracy of the Section \ref{S:stageII} asymptotic variance expression  for   the mean group differences.  
Mimicking the motivating TCGA breast cancer studies, we simulated $R=500$  independent datasets, each consisting of   $J=7$ observational studies, $K=2$  groups, and $L=1$ (i.e., univariate)  outcomes for $\tilde{N}$ subjects whose covariate vectors were sampled with replacement from the $N=450$  TCGA breast cancer patients.
We first took $\tilde{N}=500$ subjects in two simulation scenarios, labeled ``high'' and ``low,'' to represent the relative degrees of covariate similarity or  balance among the $JK=14$ study-group combinations; in other words,   the low similarity scenario represented higher   confounding levels. We then applied the   Section~\ref{S:stageII} procedure to meta-analyze the four  studies in each artificial dataset. Additionally, by increasing  $\tilde{N}$ from $125$, to $250$, and  then to $500$ subjects,  we 
compared the asymptotic and bootstrap-based variances  of   the group mean difference,  
 $(\lambda^{(1)}-\lambda^{(2)})$, where $\lambda^{(z)}=\mathbb{E}[Y^{(z)}]$.

As a common initial step  to all $500$ artificial datasets, we performed  k-means clustering of the covariates, $\mathbf{X}_1,\ldots,\mathbf{X}_N$, of the  TCGA datasets and detected lower-dimensional structure by aggregating them into $Q=10$  
clusters with centers $\mathbf{q}_1,\ldots,\mathbf{q}_Q \in \mathcal{R}^p$ and $m_1,\ldots,m_Q$ allocated number of covariates. Independently for the artificial datasets $r=1,\ldots,500$ comprising $\tilde{N}$ patients each, we generated the data as follows: 
 
\begin{enumerate}

\item \textbf{Natural population} \quad  Generate cluster relative weights, $\boldsymbol{\pi}^{(r)}=(\pi_1^{(r)},\ldots,\pi_Q^{(r)}) \sim \mathcal{D}_Q(\mathbf{1}_Q)$, denoting the Dirichlet distribution on the unit simplex $\mathcal{S}_{Q}$ and   $\mathbf{1}_{Q}$ representing the vector of $Q$ ones. Let the number of patients in the large natural population be $N_0=10^6$. For the natural population patients, sample their cluster memberships from the mixture distribution of integers: $c_{ir}^{(0)} \stackrel{\text{i.i.d.}}\sim \sum_{u=1}^Q \pi_u^{(r)}\zeta_{u}$ where $\zeta_{u}$ represents a point mass at $u$. Thence, pick covariate $\mathbf{x}_{ir}^{(0)}$ uniformly from the $m_{c_{ir}^{(0)}}$ TCGA covariates  allocated previously to the $c_{ir}^{(0)}$th k-means cluster. Generate the ``known'' relative group proportions  in the natural population: $\boldsymbol{\theta}^{(r)} \sim \mathcal{D}_K(\mathbf{1}_K)$, for $K=2$ groups. Fix the association between group memberships and covariates: $\delta^{(r)}_{z}(\mathbf{x}) \propto$ $1$ if $z=1$ and $\delta^{(r)}_{z}(\mathbf{x}) \propto \exp\bigl(\omega_0^{(r)} + \omega_1^{(r)}\sum_{t=1}^p x_t/\frac{1}{N_0}\sum_{i=1}^{N_0}\sum_{t=1}^p x_{irt}^{(0)}\bigr)$ if $z=2$. Here,  $\omega_1^{(r)}$  equals 1 and 0.1 in the high and low similarity scenarios respectively, with  $\omega_0^{(r)}$  chosen so that  $\delta^{(r)}_{z}(\mathbf{x}_{ir}^{(0)})$, averaged over the natural population, equals   $\theta_z^{(r)}$.

\smallskip

 \item \textbf{Covariates} \quad  
For subject $i=1,\ldots,\tilde{N}$,  sample   covariate vector $\tilde{\mathbf{x}}_i^{(r)}=(\tilde{x}_{i1}^{(r)},\ldots,\tilde{x}_{ip}^{(r)})' $ with replacement  from the  $N=450$ TCGA covariate vectors.  

\item\label{sz combo} \textbf{Study and group memberships} \quad Study $s_i^{(r)}$ and group $z_i^{(r)}$ were generated as follows:
          \begin{enumerate}

                \item  \textit{Multiple propensity score} \quad  Define the group-specific study propensities as follows:\\ $\log \bigl(\delta_{S=s\mid Z=z}(\mathbf{x})/\delta_{S=1\mid Z=z}(\mathbf{x})\bigr)=$ $sz\omega^{(r)}\sum_{t=1}^p\tilde{x}_{it}^{(r)}/\frac{1}{\tilde{N}}\sum_{i'=1}^{\tilde{N}}\sum_{t=1}^p\tilde{x}_{i't}^{(r)}$ for $s=2,\ldots,J$ and $z=1,2$. We set similarity parameter $\omega^{(r)}$   equal to 0.5 and 0.05  in the high and low similarity scenario, respectively. Assuming the same group PS as the natural population, the MPS is  available as $\delta_{sz}(\mathbf{x})=\delta_{s\mid z}(\mathbf{x})\delta_{z}(\mathbf{x})$. For patient $i=1,\ldots,\tilde{N}$, evaluate their probability vector $\boldsymbol{\delta}^{(r)}(\mathbf{x}_i)=$ $\bigl(\delta_{11}^{(r)}(\mathbf{x}_i),\ldots,\delta_{JK}^{(r)}(\mathbf{x}_i)\bigr)$
               
               \item \textit{Study-group memberships} \quad For patient $i=1,\ldots,\tilde{N}$, generate   $(s_i^{(r)},z_i^{(r)})$ from the categorical distribution with parameter $\boldsymbol{\delta}^{(r)}(\mathbf{x}_i)$.
            
\end{enumerate}

   \item \textbf{Subject-specific observed outcomes}\label{simulated response} \quad 
     Generate  $Y_i^{(r)} \mid Z_i=z_i^{(r)} \stackrel{\text{indep}}\sim N\bigl( z_i^{(r)}\sum_{t=1}^p\tilde{x}_{it}^{(r)}, \tau_{r}^2\bigr)$,   with $\tau_{r}^2$ chosen to achieve an approximate  $R$-squared of  $0.9$.

\end{enumerate}

Subsequently, we disregarded knowledge of all  simulation parameters and analyzed each artificial dataset using the proposed  methods.  As discussed in Section~\ref{S:stageII}, during   Stage 1 of the inferential procedure, we  estimated the  MPS  of each dataset. We then evaluated   the unnormalized balancing  weights, $\tilde{\rho}_1,\ldots,\tilde{\rho}_{\tilde{N}}$, for the IC, IGO, and FLEXOR pseudo-populations. The computational costs of evaluating the FLEXOR weights were negligible.

\begin{table}
\small
\centering
	\begin{tabular}{l|ccc|ccc}
	\toprule
		&\multicolumn{3}{c}{Low similarity}&\multicolumn{3}{c}{High similarity}\\
			\midrule
		    &\textbf{FLEXOR} &\textbf{IGO}  &\textbf{IC} &\textbf{FLEXOR} &\textbf{IGO}  &\textbf{IC} \\
	\midrule
	Minimum &78.37 &20.52 &20.61 &85.81 &30.80 &30.62\\
	First quartile &85.51 &29.91 &29.83 &94.08 &55.67 &55.61\\
	Median &87.26 &32.07 &31.92 &95.19 &73.73 &73.70\\
	Mean &87.20 &31.97 &31.87 &95.03 &70.21 &69.90\\
	Third quartile &89.02 &34.46 &34.50 &96.17 &86.09 &85.62\\
	Maximum &93.92 &42.56 &42.87 &98.59 &94.52 &94.61\\
		\bottomrule
	\end{tabular}
 \vspace{25 pt}
		\caption{For the 500 simulated datasets,  percentage ESS summaries for the three pseudo-populations in the low and high simulation scenarios with $\tilde{N}=500$ subjects.}\label{table:simulation ESS}
\end{table}

Define \textit{percent ESS}  as the ESS for 100 participants. For $\tilde{N}=500$ subjects, 
Table \ref{table:simulation ESS} presents summaries of the percent ESS of the  FLEXOR, IGO, and IC  pseudo-populations in the low and high similarity scenarios.  Unsurprisingly, all three pseudo-populations had  substantially higher ESS in the less challenging high similarity scenario in which the covariates were almost balanced   even before applying the weighting methods.   In both scenarios, the IC and IGO pseudo-populations  had  similar  ESS and   a median ESS of approximately 32\% (74\%) in  the low (high) simulation scenarios. The FLEXOR pseudo-population had  substantially higher ESS in every dataset and scenario, and median ESS   of 87.26\% (95.19\%\%)  in the low (high) scenarios   corresponding to 436.3 and 475.95 subjects, respectively.

We applied the Section \ref{S:stageII}  strategy to make weighted inferences about  functionals of the group-specific means $\lambda^{(z)}$ and standard deviations $\sigma^{(z)}$ 
of the $z$th group's potential outcomes. The  sufficient conditions of Theorem \ref{Thm:estimation} and its corollaries  are satisfied by  the potential outcome features and   pseudo-populations.  
 Since the estimands  depend on the 
pseudo-population, we evaluated each estimator's accuracy relative to the true value of its corresponding estimand computed using Monte Carlo methods.

%%%%%%%%%%%%%%%%%%%%%%%%%%%%
%%%%%%%%%%%%%%%%%%%%%%%%%%%%
%%%%%%%%%%%%%%%%%%%%%%%%%%%%

\begin{table}
\footnotesize
\centering
	\begin{tabular}{l|ccc |ccc|ccc}
	\toprule\midrule
		\multicolumn{10}{c}{\Large \textbf{Low similarity scenario} }\\
			\midrule\midrule
		     &\multicolumn{3}{c}{\large \textbf{\textit{Absolute bias}$\times 10^2$}}&\multicolumn{3}{c}{\large \textbf{\textit{Standard deviation}}$\times 10$} &\multicolumn{3}{c}{\large \textbf{\textit{Coverage} (\%)}}\\
		\midrule
		   \textbf{Estimand} &\textbf{FLEXOR} &\textbf{IGO}  &\textbf{IC} &\textbf{FLEXOR} &\textbf{IGO}  &\textbf{IC} &\textbf{FLEXOR} &\textbf{IGO}  &\textbf{IC} \\
	\midrule
$\lambda^{(1)}$ &$\bm{2.9}$ ($0.1$)  &$4.1$ ($0.1$)  &$3.8$ ($0.1$)  &$\bm{2.9}$ ($0.0$)  &$3.6$ ($0.0$)  &$3.4$ ($0.0$)  &$97$ &$93$ &$95$ \\
$\lambda^{(2)}$ &$\bm{4.5}$ ($0.1$)  &$8.4$ ($0.2$)  &$6.6$ ($0.1$)  &$\bm{4.6}$ ($0.1$)  &$6.5$ ($0.1$)  &$5.7$ ($0.1$)  &$98$ &$88$ &$89$ \\
\midrule
$\sigma^{(1)}$ &$\bm{2.6}$ ($0.1$)  &$3.0$ ($0.1$)  &$3.1$ ($0.1$)  &$2.6$ ($0.1$)  &$2.6$ ($0.0$)  &$2.8$ ($0.1$)  &$95$ &$90$ &$90$ \\
$\sigma^{(2)}$ &$\bm{4.4}$ ($0.1$)  &$6.0$ ($0.1$)  &$6.2$ ($0.1$)  &$\bm{3.8}$ ($0.1$)  &$4.5$ ($0.1$)  &$4.8$ ($0.1$)  &$93$ &$89$ &$89$ \\
\midrule
$\lambda^{(1)}-\lambda^{(2)}$ &$\bm{4.6}$ ($0.1$)  &$7.9$ ($0.2$)  &$7.4$ ($0.2$)  &$\bm{4.4}$ ($0.1$)  &$6.1$ ($0.0$)  &$6.1$ ($0.0$)  &$96$ &$89$ &$90$ \\
			\midrule
		\midrule
		\multicolumn{10}{c}{\Large \textbf{High similarity scenario} }\\
			\midrule\midrule
		     &\multicolumn{3}{c}{\large \textbf{\textit{Absolute bias}$\times 10^2$}}&\multicolumn{3}{c}{\large \textbf{\textit{Standard deviation}}$\times 10$} &\multicolumn{3}{c}{\large \textbf{\textit{Coverage} (\%)}}\\
		\midrule
$\lambda^{(1)}$ &$2.8$ ($0.1$)  &$3.3$ ($0.1$)  &$2.7$ ($0.1$)  &$2.8$ ($0.0$)  &$3.2$ ($0.0$)  &$2.8$ ($0.0$)  &$97$ &$96$ &$96$ \\
$\lambda^{(2)}$ &$4.4$ ($0.1$)  &$5.7$ ($0.1$)  &$4.2$ ($0.1$)  &$4.5$ ($0.0$)  &$5.5$ ($0.0$)  &$4.6$ ($0.0$)  &$97$ &$95$ &$97$ \\
\midrule
$\sigma^{(1)}$ &$3.0$ ($0.1$)  &$2.9$ ($0.1$)  &$2.7$ ($0.1$)  &$2.2$ ($0.0$)  &$2.3$ ($0.0$)  &$2.5$ ($0.0$)  &$94$ &$94$ &$94$ \\
$\sigma^{(2)}$ &$6.1$ ($0.2$)  &$5.9$ ($0.1$)  &$\bm{5.2}$ ($0.1$)  &$3.9$ ($0.0$)  &$4.1$ ($0.1$)  &$4.5$ ($0.1$)  &$93$ &$94$ &$94$ \\
\midrule
$\lambda^{(1)}-\lambda^{(2)}$ &$4.3$ ($0.1$)  &$4.7$ ($0.1$)  &$4.5$ ($0.1$)  &$\bm{4.1}$ ($0.0$)  &$4.4$ ($0.0$)  &$4.4$ ($0.0$)  &$97$ &$95$ &$96$ \\
			\midrule
		%%%%%%%%%%%%%%%%%%
		\bottomrule
	\end{tabular}
 \vspace{25 pt}
		\caption{In the two  simulation scenarios with $\tilde{N}=500$ subjects each,  averaging over  500 artificial datasets, the absolute biases, variances, and 95\% confidence interval coverages of some potential outcome features for the FLEXOR, IGO, and IC pseudo-populations.  For each artificial dataset and weighting method, the three performance measures were estimated using 500 independent bootstrap samples. The estimated standard errors are displayed in parentheses. For each feature (row), and separately for the absolute bias and variance measures (three-column block), a weighting strategy (column) is marked in bold if it significantly outperforms the other two strategies.}\label{table:simulation 2}
\end{table}

For various estimands and both similarity scenarios, and averaging over the 500 artificial datasets comprising $\tilde{N}=500$ subjects each, 
 Table  \ref{table:simulation 2} 
displays  the absolute biases, variances, and coverages of nominally 95\%  confidence intervals for the FLEXOR, IGO, and IC pseudo-populations.  For each artificial dataset and weighting method, the three performance measures were estimated using 500 independent bootstrap samples. For each potential outcome feature (row), and separately for the absolute bias and variance performance measures (three-column block), a pseudo-population (column) is marked in bold if it significantly outperforms the other competing pseudo-populations.
 In general, the IGO and IC weights had comparable performances for these data. The three methods had somewhat similar accuracies and reasonable coverages in the high similarity  scenario where the covariates were almost balanced across the study-group combinations. However,
in the more realistic and challenging low similarity  simulation scenario, the best results  typically corresponded to the FLEXOR pseudo-population, which often   substantially outperformed   the other methods.  Somewhat unexpectedly, this included  the mean group difference $(\lambda^{(1)}-\lambda^{(2)})$, for which  IGO weights are theoretically  optimal  under additional  assumptions such as homoscedasticity \citep[see][for single studies]{li2019propensity};  the simulation mechanism did not comply with these  sufficient conditions.    
 The results demonstrate the advantages of the FLEXOR  strategy which focuses on stabilizing the  balancing weights  rather than inferences about specific  estimands. 

Finally, we compared  the bootstrap-based and asymptotic variances of estimator (\ref{Yz}) for unconfounded  inferences about the mean group difference, $\lambda^{(1)}-\lambda^{(2)}$. For an increasing number of subjects, i.e., $\tilde{N}=125$, $250$, and $500$, we generated 500 artificial datasets in the high and low similarity scenarios. For any dataset, 
the  asymptotic variance of  weighted estimator $\hat{\lambda}^{(1)}-\hat{\lambda}^{(2)}$ is available by applying Theorem~\ref{Thm:estimation} and  the  subsequently discussed group comparison strategies. This theoretical limiting value can be 
 compared to the variance estimate   based on $B=500$ bootstrap samples. 
Supplementary Materials compares these numbers for the simulation scenarios and  sample sizes. We find that when the sample size is relatively small (i.e., $\tilde{N}\le 250$), there is a substantial difference between the asymptotic and  bootstrap-based variances. This difference indicates that a sufficiently large number of samples may be required for the asymptotic variance to be reliable.
However, for $\tilde{N}=500$ subjects, the two variances match very well, giving us the confidence to  use  asymptotic variances in the TCGA data analysis with a comparable number of patients.  

%%%%%%%%%%%%%%%%%%%%%%%%%%%%%%

%%%%%%%%%%%%%%%%%%%%%%%%%%%%%%

\section{Data Analysis} \label{S:data analysis}

To understand  breast cancer oncogenesis, 
we analyzed  the $J=7$ motivating TCGA  studies  using mRNA expression measurements on  $L=8$ targeted  genes and $p=30$ demographic and  clinicopathological covariates for  $N=450$  patients. The participants are partitioned into $K=2$ groups determined by cancer subtypes IDC and ILC,  constituting approximately 80\% and 10\% of  U.S. breast cancer cases \citep{IDC,ILC};  the study-specific percentages    in Supplementary Materials are significantly different. 
  
 The  ESS of the IC  weights was 25.7\% or  115.7 patients. The IGO weights had   a similar ESS of 26.4\% or 118.7 patients. The FLEXOR population had a  higher ESS of 40.9\% or 183.9 patients, while also guaranteeing that the  weight-adjusted composition of IDC and ILC patients in each TCGA study matched  the composition of  U.S. breast cancer patients.
Applying the Section \ref{S:stageII} procedure, we estimated population-level functionals of the group potential outcomes for the FLEXOR, IC, and IGO pseudo-populations.  For example, for the $l$th biomarker, the group-specific mean $\lambda_l^{(z)}$ and standard deviation $\sigma^{(z)}$ were estimated  by setting  $\mathbf{\Phi}(\mathbf{y})=(y_l,y_l^2)'$  in Theorem \ref{Thm:estimation}  and $\psi(t_1,t_2)=\sqrt{t_2-t_1^2}$ in Corollary~\ref{corollary 2}. Median $M_l^{(z)}$ was estimated by first estimating the CDF of potential outcome $Y^{(z)}_l$ for a fine grid of points. Group comparison estimands  like $\lambda_l^{(1)}-\lambda_l^{(2)}$ and $\sigma^{(1)}/\sigma^{(2)}$ were estimated by applying appropriately defined functionals to the estimates of $\lambda_l^{(1)}$, $\lambda_l^{(2)}$,  $\sigma^{(1)}_l$, and $\sigma^{(2)}_l$. 
 The estimate  and 95\%  confidence interval based on $B=100$ bootstrap samples are displayed in Table \ref{table:estimands1a} for each  feature (row),  pseudo-population (column), and  genes COL9A3, CXCL12, IGF1, and ITGA11 (block). The results for the genes IVL, LEF1, IC, and SMR3B are displayed in  Supplementary Materials. For each gene-estimand combination, a confidence interval for the IC or IGO pseudo-population is marked in bold  whenever the FLEXOR pseudo-population's confidence interval  was \textit{narrower};
we find that  the FLEXOR pseudo-population  often provided the most precise (narrowest) confidence intervals. 

For  FLEXOR,   the   confidence intervals for $\lambda_l^{(1)}-\lambda_l^{(2)}$ reveal that the  mean potential outcomes  were significantly different between the  disease subtypes for  genes CXCL12, IGF1, LEF1,  PRB2,  and SMR3B. 
Additionally, the standard deviation of the   IDC and IDL potential outcomes for FLEXOR were substantially different for the genes COL9A3, PRB2, and IVL;  the respective confidence intervals for  $\sigma_l^{(1)}/\sigma_l^{(2)}$  excluded 1. If required, the group-specific medians could  be  compared by inferences on   $M_l^{(1)}/M_l^{(2)}$ or $M_l^{(1)}-M_l^{(2)}$.

%%%%%%%%%%%%%%%%%%%%%%%%%%%%%%

\begin{table}
\scriptsize
\centering
\begin{tabular}{llll}
\toprule\midrule
\multicolumn{4}{c}{\Large \textbf{COL9A3} ($l=1$)}\\
\midrule
\textbf{Estimand} &\textbf{FLEXOR} &\textbf{IC} &\textbf{IGO}\\
\midrule
$\lambda_l^{(1)}$ &$-0.05$ ($-0.27,0.23)$  &$-0.11$ ($\bm{-0.33,0.21}$)  &$-0.09$ ($-0.25,0.22)$  \\
$\lambda_l^{(2)}$ &$-0.11$ ($-0.36,0.20)$  &$-0.16$ ($\bm{-0.44,0.21}$)  &$-0.19$ ($\bm{-0.49,0.27}$)  \\
\midrule
$\sigma_l$$^{(1)}$ &$1.03$ ($0.87,1.26)$  &$0.97$ ($\bm{0.84,1.31}$)  &$0.93$ ($\bm{0.83,1.36}$)  \\
$\sigma_l$$^{(2)}$ &$0.68$ ($0.54,0.86)$  &$0.69$ ($\bm{0.47,0.90}$)  &$0.69$ ($\bm{0.49,0.91}$)  \\
\midrule
M$_l^{(1)}$ &$-0.19$ ($-0.47,0.07)$  &$-0.23$ ($\bm{-0.46,0.13}$)  &$-0.23$ ($-0.40,0.11)$  \\
M$_l^{(2)}$ &$-0.05$ ($-0.52,0.37)$  &$0.07$ ($\bm{-0.57,0.49}$)  &$-0.05$ ($\bm{-0.54,0.48}$)  \\
\midrule
$\lambda_l^{(1)}-\lambda_l^{(2)}$ &$0.06$ ($-0.33,0.43)$  &$0.05$ ($\bm{-0.48,0.54}$)  &$0.09$ ($\bm{-0.36,0.59}$)  \\
$\sigma_l^{(1)}/\sigma_l^{(2)}$ &$1.52$ ($1.15,2.09)$  &$1.41$ ($\bm{1.11,2.24}$)  &$1.35$ ($\bm{1.09,2.24}$)  \\
\midrule
\multicolumn{4}{c}{\Large \textbf{CXCL12} ($l=2$)}\\
\midrule
\textbf{Estimand} &\textbf{FLEXOR} &\textbf{IC} &\textbf{IGO}\\
\midrule
$\lambda_l^{(1)}$ &$-0.03$ ($-0.22,0.21)$  &$-0.03$ ($\bm{-0.23,0.29}$)  &$0.02$ ($\bm{-0.31,0.22}$)  \\
$\lambda_l^{(2)}$ &$0.59$ ($0.23,0.88)$  &$0.55$ ($\bm{0.11,1.01}$)  &$0.58$ ($\bm{0.26,1.01}$)  \\
\midrule
$\sigma_l$$^{(1)}$ &$0.91$ ($0.84,1.16)$  &$0.97$ ($\bm{0.84,1.21}$)  &$0.94$ ($\bm{0.83,1.21}$)  \\
$\sigma_l$$^{(2)}$ &$0.80$ ($0.52,1.10)$  &$0.83$ ($\bm{0.49,1.27}$)  &$0.82$ ($\bm{0.54,1.20}$)  \\
\midrule
M$_l^{(1)}$ &$-0.15$ ($-0.20,0.36)$  &$-0.16$ ($\bm{-0.32,0.36}$)  &$-0.09$ ($\bm{-0.35,0.38}$)  \\
M$_l^{(2)}$ &$0.68$ ($0.44,1.01)$  &$0.69$ ($\bm{0.10,1.16}$)  &$0.58$ ($\bm{0.21,1.08}$)  \\
\midrule
$\lambda_l^{(1)}-\lambda_l^{(2)}$ &$-0.62$ ($-1.00,-0.12)$  &$-0.58$ ($\bm{-1.18,-0.08}$)  &$-0.56$ ($\bm{-1.08,-0.17}$)  \\
$\sigma_l^{(1)}/\sigma_l^{(2)}$ &$1.14$ ($0.87,1.99)$  &$1.17$ ($\bm{0.71,2.32}$)  &$1.14$ ($\bm{0.75,1.94}$)  \\
\midrule
\multicolumn{4}{c}{\Large \textbf{IGF1} ($l=3$)}\\
\midrule
\textbf{Estimand} &\textbf{FLEXOR} &\textbf{IC} &\textbf{IGO}\\
\midrule
$\lambda_l^{(1)}$ &$0.04$ ($-0.21,0.23)$  &$0.10$ ($\bm{-0.28,0.31}$)  &$0.13$ ($\bm{-0.30,0.30}$)  \\
$\lambda_l^{(2)}$ &$0.82$ ($0.54,1.09)$  &$0.84$ ($\bm{0.52,1.16}$)  &$0.82$ ($\bm{0.52,1.17}$)  \\
\midrule
$\sigma_l$$^{(1)}$ &$0.81$ ($0.80,1.12)$  &$0.86$ ($\bm{0.78,1.18}$)  &$0.87$ ($\bm{0.75,1.16}$)  \\
$\sigma_l$$^{(2)}$ &$0.76$ ($0.47,0.95)$  &$0.82$ ($\bm{0.45,1.13}$)  &$0.76$ ($\bm{0.45,1.02}$)  \\
\midrule
M$_l^{(1)}$ &$-0.01$ ($-0.18,0.34)$  &$0.06$ ($\bm{-0.23,0.47}$)  &$0.10$ ($\bm{-0.26,0.45}$)  \\
M$_l^{(2)}$ &$0.95$ ($0.60,1.22)$  &$0.95$ ($\bm{0.43,1.28}$)  &$0.88$ ($\bm{0.52,1.32}$)  \\
\midrule
$\lambda_l^{(1)}-\lambda_l^{(2)}$ &$-0.77$ ($-1.22,-0.43)$  &$-0.74$ ($\bm{-1.19,-0.33}$)  &$-0.69$ ($\bm{-1.18,-0.31}$)  \\
$\sigma_l^{(1)}/\sigma_l^{(2)}$ &$1.06$ ($0.94,2.03)$  &$1.05$ ($\bm{0.83,2.33}$)  &$1.14$ ($\bm{0.82,2.26}$)  \\
\midrule
\multicolumn{4}{c}{\Large \textbf{ITGA11} ($l=4$)}\\
\midrule
\textbf{Estimand} &\textbf{FLEXOR} &\textbf{IC} &\textbf{IGO}\\
\midrule
$\lambda_l^{(1)}$ &$0.01$ ($-0.28,0.22)$  &$0.03$ ($\bm{-0.37,0.24}$)  &$0.07$ ($-0.29,0.17)$  \\
$\lambda_l^{(2)}$ &$0.01$ ($-0.48,0.26)$  &$-0.02$ ($\bm{-0.53,0.27}$)  &$0.07$ ($\bm{-0.63,0.28}$)  \\
\midrule
$\sigma_l$$^{(1)}$ &$0.92$ ($0.83,1.10)$  &$0.96$ ($\bm{0.80,1.16}$)  &$0.94$ ($\bm{0.83,1.19}$)  \\
$\sigma_l$$^{(2)}$ &$0.81$ ($0.60,1.03)$  &$0.93$ ($\bm{0.54,1.07}$)  &$0.98$ ($\bm{0.56,1.15}$)  \\
\midrule
M$_l^{(1)}$ &$0.14$ ($-0.28,0.41)$  &$0.19$ ($\bm{-0.49,0.48}$)  &$0.20$ ($\bm{-0.36,0.39}$)  \\
M$_l^{(2)}$ &$-0.02$ ($-0.54,0.32)$  &$-0.22$ ($\bm{-0.72,0.26}$)  &$-0.09$ ($\bm{-0.55,0.35}$)  \\
\midrule
$\lambda_l^{(1)}-\lambda_l^{(2)}$ &$0.01$ ($-0.28,0.49)$  &$0.05$ ($\bm{-0.41,0.56}$)  &$0.00$ ($\bm{-0.44,0.60}$)  \\
$\sigma_l^{(1)}/\sigma_l^{(2)}$ &$1.14$ ($0.89,1.62)$  &$1.03$ ($\bm{0.86,2.01}$)  &$0.96$ ($\bm{0.82,1.84}$)  \\
\midrule
\bottomrule
\end{tabular}
		\caption{For four targeted genes, estimates and 95\% bootstrap confidence levels (shown in parentheses) of different population-level estimands of the  potential outcomes of group 1 (IDC cancer subtype, denoted by superscript 1) and group 2 (ILC cancer subtype, denoted by superscript 2)
with FLEXOR, IC, and IGO weights.  An IC or IGO confidence interval is highlighted in bold  if it is \textit{wider} than the  FLEXOR confidence
interval. All numbers are rounded to 2 decimal places. See Section \ref{S:data analysis} for further explanation.}\label{table:estimands1a}
\end{table}

%%%%%%%%%%%%%%%%%%%%%%%%%%%%%%

Next, we estimated
the correlation between the potential outcomes of the $l_1$th and $l_2$th biomarker in the $z$th group: for  $M=3$ and $\mathbf{y} \in \mathcal{R}^8$, we assumed an $M$-variate function, $\mathbf{\Phi}(\mathbf{y})=(\Phi_1(\mathbf{y}), \Phi_2(\mathbf{y}), \Phi_3(\mathbf{y}))'$, with component functions,  $\Phi_1(\mathbf{y})=y_{l_1}$, $\Phi_2(\mathbf{y})=y_{l_3}$, and $\Phi_3(\mathbf{y})=y_{l_1}y_{l_2}$. For the $z$th group, we  estimated  $\boldsymbol{\lambda}^{(z)}=\bigl(\mathbb{E}[Y_{l_1}^{(z)}],\, \mathbb{E}[Y_{l_2}^{(z)}],\, \mathbb{E}[Y_{l_1}^{(z)}Y_{l_2}^{(z)}]\bigr)'$ for a pseudo-population by applying Theorem \ref{Thm:estimation}.  Setting $\psi(t_1,t_2,t_3)=t_3-t_1t_2$, we then applied Corollary \ref{corollary 2} to  estimate  pseudo-population covariance, $\psi(\boldsymbol{\lambda}^{(z)})=\text{cov}(Y_{l_1}^{(z)},Y_{l_2}^{(z)})$. Using the  estimated standard deviations $\sigma_{l_1}^{(z)}$ and $\sigma_{l_2}^{(z)}$ for the  pseudo-population, as described above, we  estimated  the  correlation. Independent estimates from $B=100$ bootstrap samples were used to compute 95\% confidence intervals of  the true correlation between the $l_1$th and $l_2$th gene pair in the $z$th group. 
Supplementary Materials present   95\% confidence intervals of  the  group-specific correlations for each gene pair and weighting method.

%%%%%%%%%%%%%%%%%%%%%%%%%%%%%%
\begin{table}
\centering
	\begin{tabular}{ll}
\toprule\midrule
\multicolumn{2}{c}{\large \textbf{Infiltrating Ductal Carcinoma}}\\
\midrule
\textbf{Pseudo-population} &\textbf{Significantly correlated gene pairs} \\
\midrule
FLEXOR &CXCL12-IGF1, CXCL12-ITGA11, IGF1-ITGA11,  \\
&COL9A3-LEF1, CXCL12-LEF1, IGF1-LEF1,  \\
&ITGA11-LEF1, IVL-LEF1, COL9A3-PRB2 \\
\\
IC &CXCL12-IGF1, CXCL12-ITGA11, IGF1-ITGA11,  \\
&CXCL12-LEF1, IGF1-LEF1, COL9A3-PRB2 \\
\\
IGO &CXCL12-IGF1, CXCL12-ITGA11, IGF1-ITGA11,  \\
&CXCL12-LEF1, IGF1-LEF1, COL9A3-PRB2 \\
\midrule
\multicolumn{2}{c}{\large \textbf{Infiltrating Lobular Carcinoma}}\\
\midrule
\textbf{Pseudo-population} &\textbf{Significantly correlated gene pairs} \\
\midrule
FLEXOR &CXCL12-IGF1 \\
\\
IC &CXCL12-IGF1 \\
\\
IGO &CXCL12-IGF1 \\
\midrule
\bottomrule
\end{tabular}
\vspace{25 pt}
\caption{Co-expressed gene pairs for each  pseudo-population and  breast cancer subtype.}\label{table:pairs}
\end{table}
 %%%%%%%%%%%%%%%%%%%%%%%%%%%%%%%%%%%%%%%%%%%%%%%%%%
 
 Table \ref{table:pairs} lists the significantly correlated gene pairs for each disease subtype. 
For the  FLEXOR pseudo-population and IDC disease subtype, gene CXCL12 was significantly co-expressed with  the   IGF1, ITGA11, and LEF1; gene  IGF1 was co-expressed with  ITGA11 and LEF1; gene COL9A3 was co-expressed with   LEF1 and PRB2; and gene LEF1 was co-expressed with   IVL and ITGA11. For  disease subtype ILC, only the CXCL12 - IGF1 gene pair was significantly correlated according to FLEXOR. The {differential correlation pattern} for the FLEXOR pseudo-population was, therefore, the gene pairs (CXCL12, ITGA11), (IGF1, ITGA11),  (COL9A3, LEF1), (CXCL12, LEF1), (IGF1, LEF1),  (ITGA11, LEF1), (IVL, LEF1), and (COL9A3, PRB2).  Detecting these variations in gene co-expression patterns between the IDC and ILC subtypes of breast cancer patients in the United States is crucial for informing precision medicine and targeted therapies  \citep{schmidt2016precision}.

\begin{figure}
\centering
\includegraphics[scale=.4]{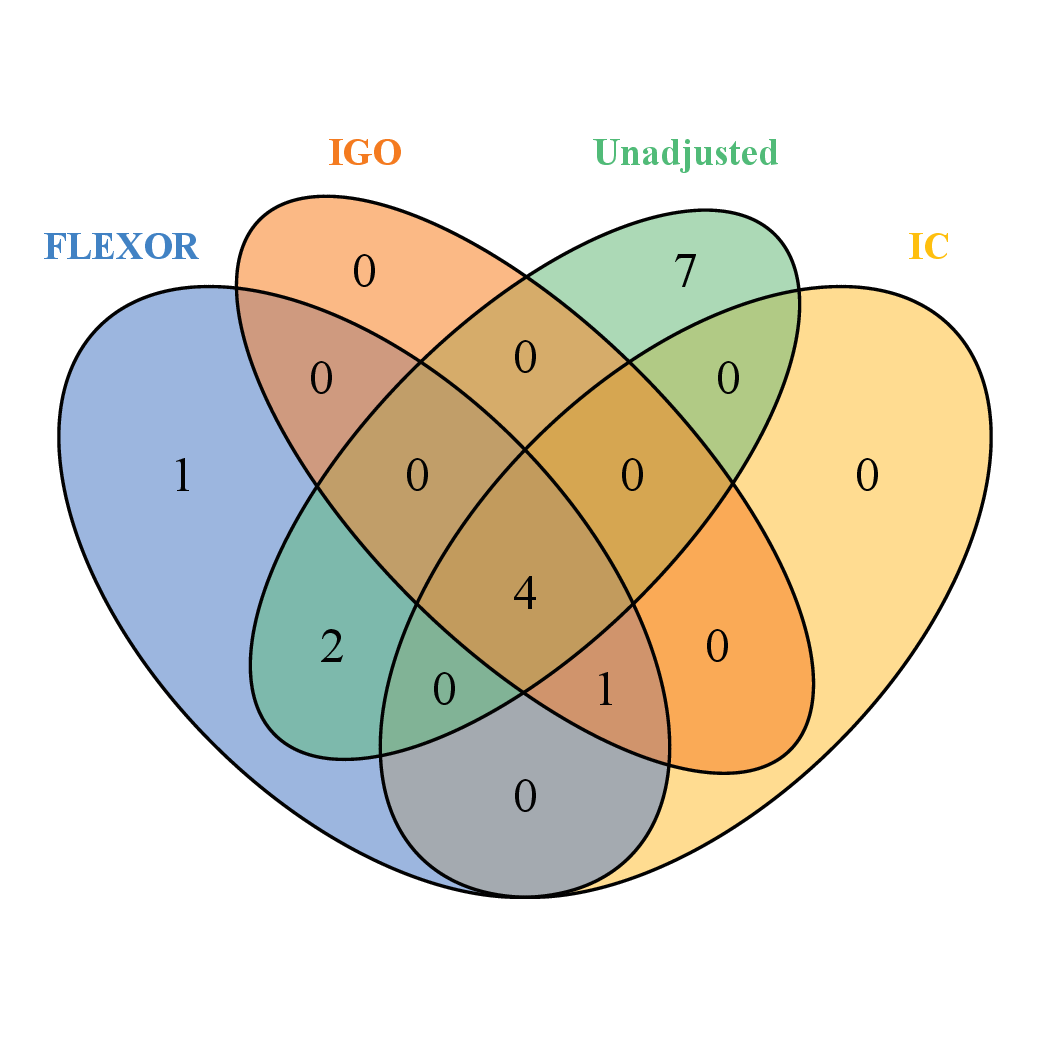}
\caption{Venn diagram of the   differential correlation pattern of the targeted gene pairs for the  three weighting methods and  unweighted analysis.
}\label{F:venn}
\end{figure}

By contrast,   Table \ref{table:pairs} shows that
the differential correlation pattern of the IC pseudo-population comprised just five gene pairs, and was identical to the IGO pseudo-population's pattern. Although these  gene pairs  were also detected by the FLEXOR pseudo-population, the latter  detected  additional  co-expressed gene pairs. Figure \ref{F:venn}   graphically summarizes the  number of differentially correlated  gene pairs  discovered by the  weighting methods and (biased) unadjusted analyses.   
 Recent literature on breast cancer gene ontology substantiates the distinctive findings of  FLEXOR. The genes IVL and LEF1  are highly expressed in basal and metaplastic human breast cancers, and   the cell adhesion and  ECM receptor pathways, containing the  genes ITGA11 and LEF1, are deregulated   \citep{williams2022elevated}.
 The focal adhesion and cell cycle pathways,    containing the genes COL9A3 and LEF1, are affected by    WNT signaling gene set mutations caused by breast cancer metastases \citep{paul2020genomic}.

 %%%%%%%%%%%%%%%%%%%%%%%%%%%%%%%%%%%%%%%%%%%%%%%%%%

\section{Conclusion} \label{S:discussion}

 In multiple retrospective cohorts, the integrative analysis of   mixed-type multivariate  outcomes  to accurately compare  multiple groups is a  challenging problem.     We 
 formulate new frameworks for covariate-balanced pseudo-populations that extend  existing weighting methods to  meta-analytical investigations and design a novel,  estimand-agnostic FLEXOR  pseudo-population that maximizes the effective sample  size by a cost-effective iterative procedure. We propose generally applicable  weighted estimators for a wide variety  of population-level univariate or  multivariate features  relevant to multigroup comparisons, e.g., correlation coefficients and contrasts and ratios of means, medians, and  standard deviations.

The methodology has a  wide range of meta‐analytical applications,    including  multi‐arm RCTs.  A component of the multi-study balancing weights is considerably simplified if the $s^*$th study is an RCT, in which case the  study-specific group MPS $\delta_{z|s^*}(\mathbf{x})$ equals $1/K$. In general, the  theoretical results hold for a mix of observational studies and RCTs, although the study MPS must still be estimated because  the subject-study allocations are usually non-random for multiple studies. 
 The  methodology  may  be generalized in several other directions,    such as increased efficiency by adding an outcome modeling
component \citep{mao2019propensity,zeng2023propensity}; 
transportability  \citep{westreich2017transportability} and data-fusion \citep{bareinboim2016causal,dahabreh2020toward, dahabreh2023efficient} problems, which incorporate additional information in the form of  random samples  from the natural population; and flexible machine learning  for MPS estimation that achieves 
$\sqrt{N}$ inference   \citep{chernozhukov2018double}. In recent years, weighting approaches are also challenged and  rendered ineffectual by high-dimensional genetic or genomic measurements.  Our future research will explore these  avenues.

%\setstretch{1.39}

\bibliographystyle{agsm}

 \newcommand{\noop}[1]{}

\end{document}